\documentstyle[amssymb,floats,aps,prb,epsf]{revtex}
\epsfclipon
\begin{document}
\twocolumn[\hsize\textwidth\columnwidth\hsize\csname
@twocolumnfalse\endcsname

\draft

\title{Charge transport of electron doped Mott insulators on a
triangular lattice}
\author{Bin Liu, Ying Liang, and Shiping Feng$^{*}$}
\address{Department of Physics, Beijing Normal University, Beijing
100875, China}
\author{Wei Yeu Chen}
\address{Department of Physics, Tamkang University, Tamsui 25137,
Taiwan}

\maketitle
\begin{abstract}
The charge transport of electron doped Mott insulators on a
triangular lattice is investigated within the t-J model. The
conductivity spectrum shows a low-energy peak and an unusual
midinfrared band, while the resistivity is characterized by a
crossover from the high temperature metallic-like to low
temperature insulating-like behavior, in qualitative agreement
with experiments. Our results also show that the mechanism that
causes this unusual charge transport is closely related to a
competition between the kinetic energy and magnetic energy in the
system.

\end{abstract}

]
\bigskip
\narrowtext

The recent discovery of superconductivity in doped cobaltates,
Na$_{x}$CoO$_{2}\cdot y$H$_{2}$O with the superconducting
transition temperature $T_{c}\sim 5$K, has generated great
interests due to the role of strong electron correlations
\cite{takada}. This compound has a lamellar structure consisting
of the two-dimensional (2D) CoO$_{2}$ layers separated by a thick
insulating layer of Na$^{+}$ ions and H$_{2}$O molecules, where
the one-half spin Co$^{4+}$ ions are arranged on a triangular
lattice. This structure is similar to cuprates superconductors in
the sense that they also have a layered structure of the square
lattice of the CuO$_{2}$ plane separated by insulating layers
\cite{kastner}. It has been well established that Cu$^{2+}$ ions
exhibit an antiferromagnetic (AF) long-range order (AFLRO) in the
parent compounds of cuprate superconductors, where
superconductivity occurs when the AFLRO state is suppressed by
hole or electron doping \cite{kastner,tokura}. However,
Na$_{x}$CoO$_{2}\cdot y$H$_{2}$O is viewed as an electron doped
Mott insulator, where superconductivity appears with electron
doping \cite{takada}. A fundamental similarity between cuprate and
cobaltate superconductors has been seen in the decreases in the
superconducting transition temperature for both underdoped and
overdoped materials \cite{schaak}. The optimal doping for
superconductivity occurs at $0.3$ electrons per Co above the
ground-state Na$_{0}$CoO$_{2}\cdot 1.3$H$_{2}$O, which is a
half-filled two-electron $t_{2g}$ derived band, while for
cuprates, the optimal doping occurs at $0.15$ holes (electrons)
added to the half-filled band of the parent compound
\cite{schaak}. On the other hand, doped Mott insulators on a
triangular lattice, where the geometric spin frustration exists,
are also of interests in their own right, with many unanswered
fascinating questions. Historically the undoped Mott insulator on
a triangular lattice was firstly proposed to be a model for
microscopic realization of the spin liquid due to the existence of
the strong spin frustration \cite{anderson1}. It has been argued
that this spin liquid state may play a crucial role in the
mechanism for superconductivity in doped cuprates
\cite{anderson2}. Thus the unexpected finding of superconductivity
in doped cobaltates has raised the hope that it may help solve the
unusual physics in doped cuprates.

Some experimental studies have revealed a non-Fermi liquid
behavior in doped cobaltates \cite{takada,wang,chou}. Among the
striking features in the normal-state, a hallmark is the charge
transport, where the resistivity shows a temperature linear
variation in a wide range of temperatures \cite{takada,wang}.
Moreover, this unusual charge transport is found to be
intriguingly related to the magnetic correlation. The charge
transport measurement is a powerful probe for interacting electron
systems \cite{tanner}, and can provide very detailed knowledges
about the low-energy excitations as electrons are doped to
cobaltates. One of the important issues in experimental and
theoretical investigations of doped cobaltates is understanding of
the low-energy excitations in these compounds, where the clue to
their superconductivity lies possibly already in their
nonconventional normal-state properties. Therefore it is of
interest to have a detailed look at the charge transport. In this
paper, we try to study this issue. We show that the resistivity of
electron doped cobaltates is characterized by a crossover from the
high temperature metallic-like to low temperature insulating-like
behavior, and the mechanism that causes this unusual transport is
closely related to a competition between the kinetic energy and
magnetic energy in the system.

In electron doped cobaltates, the characteristic feature is the
presence of the 2D CoO$_{2}$ plane \cite{takada} as mentioned
above, and it seems evident that the unusual behaviors are
dominated by this plane. It has been argued that the essential
physics of the doped CoO$_{2}$ plane is contained in the $t$-$J$
model on a triangular lattice \cite{baskaran},
\begin{eqnarray}
H&=&-t_{e}\sum_{i\hat{\eta}\sigma}PC_{i\sigma}^{\dagger}
C_{i+\hat{\eta}\sigma}P^{\dagger}-\mu\sum_{i\sigma}P C_{i\sigma
}^{\dagger}C_{i\sigma }P^{\dagger} \nonumber\\
&+& J\sum_{i\hat{\eta}} {\bf S}_{i}\cdot{\bf S}_{i+\hat{\eta}},
\end{eqnarray}
where $t_{e}>0$, the summation is over all sites $i$, and for each
$i$, over its nearest-neighbor $\hat{\eta}$,
$C^{\dagger}_{i\sigma}$ ($C_{i\sigma}$) is the electron creation
(annihilation) operator, ${\bf S}_{i}=C^{\dagger}_{i}{\bf
\sigma}C_{i}/2$ is the spin operator with ${\bf
\sigma}=(\sigma_{x},\sigma_{y},\sigma_{z})$ as the Pauli matrices,
$\mu$ is the chemical potential, and the projection operator $P$
removes zero occupancy, i.e., $n_{i\uparrow}+n_{i\downarrow}\geq
1$ with $n_{i\sigma}=C^{\dagger}_{i\sigma}C_{i\sigma}$. In the
past fifteen years, some useful methods have been proposed to
treat the no double occupancy local constraint in hole doped
cuprates \cite{bedell}. In particular, a fermion-spin theory based
on the partial charge-spin separation has been developed to study
the physical properties of doped cuprates \cite{feng1}, where the
no double occupancy local constraint can be treated properly in
analytical calculations. Within this theory, the charge transport
of the underdoped cuprates has been discussed \cite{feng1,feng2},
and the results are consistent with experiments
\cite{tanner,ando}. To apply this theory in electron doped
cobaltates, the $t$-$J$ model (1) can be rewritten in terms of a
particle-hole transformation $C_{i\sigma}\rightarrow
f^{\dagger}_{i-\sigma}$ as \cite{baskaran},
\begin{eqnarray}
H=-t\sum_{i\hat{\eta}\sigma}f_{i\sigma}^{\dagger}
f_{i+\hat{\eta}\sigma}+\mu\sum_{i\sigma}f_{i\sigma }^{\dagger}
f_{i\sigma }+J\sum_{i\hat{\eta}}{\bf S}_{i}
\cdot{\bf S}_{i+\hat{\eta}},
\end{eqnarray}
supplemented by a local constraint
$\sum_{\sigma}f^{\dagger}_{i\sigma}f_{i\sigma}\leq 1$ to remove
double occupancy, where as a matter of convenience, we have set
$t=-t_{e}$, $f^{\dagger}_{i\sigma}$ ($f_{i\sigma}$) is the hole
creation (annihilation) operator, and ${\bf S}_{i}=f^{\dagger}_{i}
{\bf \sigma}f_{i}/2$ is the spin operator in the hole
representation. Now we follow the partial charge-spin separation
fermion-spin theory \cite{feng1}, and decouple hole operators
$f_{i\uparrow}$ and $f_{i\downarrow}$ as,
\begin{eqnarray}
f_{i\uparrow}=a^{\dagger}_{i\uparrow}S^{-}_{i},~~~~
f_{i\downarrow}=a^{\dagger}_{i\downarrow}S^{+}_{i},
\end{eqnarray}
where the spinful fermion operator $a_{i\sigma}=
e^{-i\Phi_{i\sigma}}a_{i}$ describes the charge degree of freedom
together with the phase part of the spin degree of freedom
(dressed fermion), while the spin operator $S_{i}$ describes the
amplitude part of the spin degree of freedom (dressed spinon),
then the no double occupancy local constraint, $\sum_{\sigma}
f^{\dagger}_{i\sigma}f_{i\sigma}=S^{+}_{i}a_{i\uparrow}
a^{\dagger}_{i\uparrow}S^{-}_{i}+S^{-}_{i}a_{i\downarrow}
a^{\dagger}_{i\downarrow}S^{+}_{i}=a_{i}a^{\dagger}_{i}(S^{+}_{i}
S^{-}_{i}+S^{-}_{i}S^{+}_{i})=1-a^{\dagger}_{i}a_{i}\leq 1$, is
satisfied in analytical calculations, and the double dressed
fermion occupancy, $a^{\dagger}_{i\sigma}a^{\dagger}_{i-\sigma}=
e^{i\Phi_{i\sigma}}a^{\dagger}_{i}a^{\dagger}_{i}
e^{i\Phi_{i-\sigma}}=0$ and $a_{i\sigma}a_{i-\sigma}=
e^{-i\Phi_{i\sigma}}a_{i}a_{i}e^{-i\Phi_{i-\sigma}}=0$, are ruled
out automatically. It has been shown that these dressed fermion
and spinon are gauge invariant, and in this sense, they are real
and can be interpreted as the physical excitations. Since the
phase factor $e^{-i\Phi_{i\sigma}}$ is separated from the bare
spinon operator, therefore it describes a spinon cloud
\cite{feng1}. In this case, the dressed fermion $a_{i\sigma}$ is a
spinless fermion $a_{i}$ incorporated a spin cloud
$e^{-i\Phi_{i\sigma}}$ (magnetic flux), and is a magnetic
dressing. In other words, the gauge invariant dressed fermion
carries some spin messages, i.e., it shares some effects of the
spin configuration rearrangements due to the presence of the
electron itself \cite{martins}. Although in common sense
$a_{i\sigma}$ is not a real spinful fermion, it behaves like a
spinful fermion. In this partial charge-spin separation
fermion-spin representation, the low-energy behavior of the
$t$-$J$ model (2) can be expressed as \cite{feng1},
\begin{eqnarray}
H&=&-t\sum_{i\hat{\eta}}(a_{i\uparrow}S^{+}_{i}
a^{\dagger}_{i+\hat{\eta}\uparrow}S^{-}_{i+\hat{\eta}}+
a_{i\downarrow}S^{-}_{i}a^{\dagger}_{i+\hat{\eta}\downarrow}
S^{+}_{i+\hat{\eta}})\nonumber\\
&-&\mu\sum_{i\sigma}a^{\dagger}_{i\sigma} a_{i\sigma}+J_{{\rm
eff}}\sum_{i\hat{\eta}}{\bf S}_{i}\cdot {\bf S}_{i+\hat{\eta}},
\end{eqnarray}
with $J_{{\rm eff}}=(1-\delta)^{2}J$, and
$\delta=\langle a^{\dagger}_{i\sigma}a_{i\sigma}\rangle=\langle
a^{\dagger}_{i}a_{i}\rangle$ is the electron doping concentration.
At the zero doping, the $t$-$J$ model is reduced to the Heisenberg
model. Many authors \cite{caprioti} have shown unambiguously that
as in a square lattice, there is indeed AFLRO in the ground state
of the AF Heisenberg model on a triangular lattice. However, this
AFLRO is destroyed more rapidly with increasing doping than that
on a square lattice due to the strong geometry frustration.
Therefore there is no AFLRO away from the zero doping, i.e.,
$\langle S_{i}^{z}\rangle=0$. On the other hand, in the electron
underdoped regime where superconductivity occurs, the weak
ferromagnetism can be induced \cite{anderson3,baskaran}, since the
effective magnetic coupling in the present case is
$J_{{\rm eff}}-2\phi\mid t\mid$ with the dressed fermion mean-field
(MF) particle-hole parameter
$\phi=\langle a_{i\sigma}^{\dagger}a_{i+\hat{\eta}\sigma}\rangle$.

Since the no double occupancy local constraint has been treated
properly within the partial charge-spin separation fermion-spin
theory, this leads to disappearing of the extra gauge degree of
freedom related to the local constraint \cite{feng1}. In this
case, the charge fluctuation couples only to dressed fermions
\cite{feng1,feng2}. Based on this theory, the charge transport of
the hole doped Mott insulator on a square lattice has been
discussed \cite{feng1,feng2}, and the results are consistent with
experiments \cite{tanner,ando}. Following their discussions, the
optical conductivity of electron doped Mott insulators on a
triangular lattice can be obtained as,
\begin{eqnarray}
\sigma(\omega)&=&(Z\chi te)^{2}{1\over 2N}\sum_{k\sigma}
\gamma_{sk}^{2}\int^{\infty}_{-\infty}{d\omega'\over 2\pi}
A^{(a)}_{\sigma}({\bf k},\omega'+\omega)\nonumber \\
&\times& A^{(a)}_{\sigma}({\bf k},
\omega'){n_{F}(\omega'+\omega)-n_{F}(\omega') \over \omega},
\end{eqnarray}
where $Z$ is the number of the nearest neighbor sites, the spinon
correlation function $\chi=\langle S_{i}^{+}S_{i+\hat{\eta}}^{-}
\rangle$, $\gamma_{sk}^{2}=\{[{\rm sin}k_{x}+{\rm sin}(k_{x}/2)
{\rm cos}({\sqrt 3}k_{y}/2)]^{2}+3[{\rm sin}({\sqrt 3}k_{y}/2)
{\rm cos}(k_{x}/2)]^{2}\}/9$, $n_{F}(\omega)$ is the fermion
distribution function, and the dressed fermion spectral function
$A^{(a)}_{\sigma}(k,\omega)$ is obtained as $A^{(a)}_{\sigma}
(k,\omega)=-2{\rm Im}g_{\sigma}(k,\omega)$. The full dressed
fermion Green's function $g^{-1}_{\sigma}(k,\omega)=
g^{(0)-1}_{\sigma}(k,\omega)-\Sigma^{(a)}(k,\omega)$ with the MF
dressed fermion Green's function $g^{(0)-1}_{\sigma}(k,\omega)=
\omega-\xi_{k}$, and the second-order dressed fermion self-energy
from the dressed spinon pair bubble \cite{feng1,feng2},
\begin{eqnarray}
\Sigma^{(a)}(k,\omega)&=&{1\over 2}(Zt)^{2}{1\over N^{2}}\sum_{pq}
{B_{q+p}B_{q}\over 4\omega_{q+p}\omega_{q}}(\gamma_{q+p+k}+
\gamma_{q-k})^{2} \nonumber \\
&\times&\left ({F^{(1)}(k,p,q)\over \omega+\omega_{q+p}-
\omega_{q}-\xi_{p+k}}\right .\nonumber \\
&+&{F^{(2)}(k,p,q)\over \omega-\omega_{q+p}+
\omega_{q}-\xi_{p+k}}\nonumber \\
&+&{F^{(3)}(k,p,q)\over \omega+\omega_{q+p}+\omega_{q} -\xi_{p+k}}
\nonumber \\
&+&\left .{F^{(4)}(k,p,q)\over \omega-\omega_{q+p}-\omega_{q}-
\xi_{p+k}}\right ),
\end{eqnarray}
where $B_{k}=\lambda[(2\epsilon\chi^{z}+\chi)\gamma_{k}-(\epsilon
\chi+2\chi^{z})]$, $\lambda=2ZJ_{{\rm eff}}$,
$\epsilon=1+2t\phi/J_{{\rm eff}}$, $\gamma_{k}=[\cos{k_{x}}+
2\cos{(k_{x}/2)}\cos{({\sqrt 3}k_{y}/2)}]/3$, and
\begin{mathletters}
\begin{eqnarray}
F^{(1)}(k,p,q)&=& n_{F}(\xi_{p+k})[n_{B}(\omega_{q})-n_{B}
(\omega_{q+p})]\nonumber \\
&+&n_{B}(\omega_{q+p})[1+n_{B}(\omega_{q})], \\
F^{(2)}(k,p,q)&=& n_{F}(\xi_{p+k})[n_{B}(\omega_{q+p})-n_{B}
(\omega_{q})]\nonumber \\
&+&n_{B}(\omega_{q})[1+n_{B}(\omega_{q+p})], \\
F^{(3)}(k,p,q)&=& n_{F}(\xi_{p+k})[1+n_{B}(\omega_{q+p})+n_{B}
(\omega_{q})]\nonumber \\
&+&n_{B}(\omega_{q})n_{B}(\omega_{q+p}), \\
F^{(4)}(k,p,q)&=& [1+n_{B}(\omega_{q})][1+n_{B}(\omega_{q+p})]
\nonumber \\
&-&n_{F} (\xi_{p+k})[1+n_{B}(\omega_{q+p})+n_{B}(\omega_{q})],
\end{eqnarray}
\end{mathletters}
with $n_{B}(\omega)$ is the Bose distribution function, and the MF
dressed fermion and spinon excitation spectra are given by,
\begin{mathletters}
\begin{eqnarray}
\xi_{k}&=& Zt\chi\gamma_{k}-\mu, \\
\omega_{k}^{2}&=& A_{1} \gamma_{k}^{2}+A_{2}\gamma_{k}+A_{3},
\end{eqnarray}
\end{mathletters}
respectively, where
\begin{mathletters}
\begin{eqnarray}
A_{1}&=&\alpha\epsilon\lambda^{2}({1\over 2}\chi+
\epsilon\chi^{z}), \\
A_{2}&=&\epsilon\lambda^{2}[{1\over Z}(1-Z)\alpha({1\over 2}
\epsilon\chi+\chi^{z})\nonumber \\
&-&\alpha(C^{z}+{1\over 2}C)-{1\over 2Z}(1-\alpha)], \\
A_{3}&=&\lambda^{2}[\alpha(C^{z}+{1\over 2}\epsilon^{2}C) +
{1\over 4Z}(1-\alpha) (1+\epsilon^{2})\nonumber \\
&-&\alpha\epsilon {1\over Z} ({1\over 2}\chi+\epsilon\chi^{z})],
\end{eqnarray}
\end{mathletters}
with spinon correlation functions $\chi^{z}=\langle
S_{i}^{z}S_{i+\hat{\eta}}^{z}\rangle$,
$C=(1/Z^{2})\sum_{\hat{\eta}\hat{\eta^{\prime}}}\langle
S_{i+\hat{\eta}}^{+}S_{i+\hat{\eta^{\prime}}}^{-}\rangle$, and
$C^{z}=(1/Z^{2})\sum_{\hat{\eta}\hat{\eta ^{\prime }}} \langle
S_{i+\hat{\eta}}^{z}S_{i+\hat{\eta^{\prime }}}^{z}\rangle$. In
order to satisfy the sum rule for the correlation function
$\langle S_{i}^{+}S_{i}^{-}\rangle=1/2$ in the absence of AFLRO, a
decoupling parameter $\alpha$ has been introduced in the MF
calculation, which can be regarded as the vertex correction
\cite{feng3,kondo}. All these mean-field order parameters,
decoupling parameter, and the chemical potential are determined
self-consistently \cite{feng3}.

\begin{figure}[prb]
\epsfxsize=3.0in\centerline{\epsffile{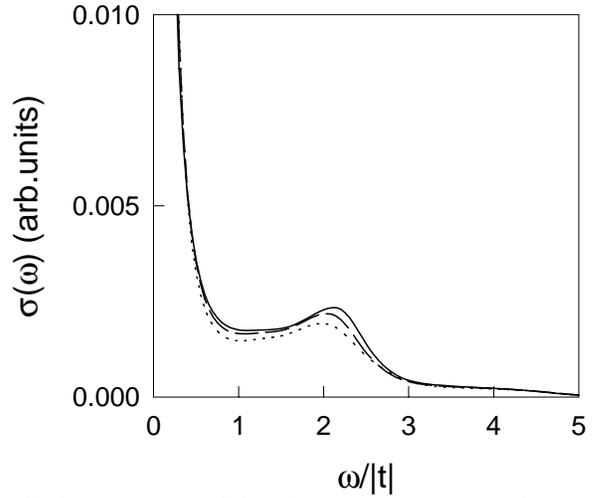}} \caption{The
conductivity of electron doped Mott insulators on a triangular
lattice as a function of frequency at electron doping
$\delta=0.40$ (solid line), $\delta=0.35$ (dashed line), and
$\delta=0.30$ (dotted line) for $t/J=-2.5$ with temperature
$T=0$.}
\end{figure}

\begin{figure}[prb]
\epsfxsize=3.0in\centerline{\epsffile{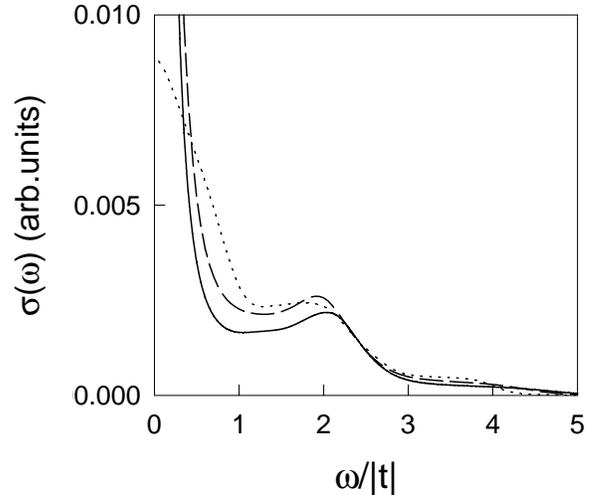}} \caption{The
conductivity of electron doped Mott insulators on a triangular
lattice as a function of frequency at electron doping
$\delta=0.35$ for $t/J=-2.5$ with temperature $T=0$ (solid line),
$T=0.3J$ (dashed line), and $T=0.7J$ (dotted line).}
\end{figure}

We have performed a numerical calculation for the optical
conductivity in Eq. (5), and the results at electron doping
$\delta=0.40$ (solid line), $\delta=0.35$ (dashed line), and
$\delta=0.30$ (dotted line) for parameter $t/J=-2.5$ in
temperature $T=0$ are shown in Fig. 1, where the charge $e$ has
been set as the unit. The conductivity spectrum shows a low-energy
peak appearing at $\omega\sim 0$, which decays rapidly, and a
broad midinfrared band. This midinfrared band is electron doping
dependent, and the component increases with increasing electron
doping for $1\mid t\mid<\omega <3\mid t\mid$, and is nearly
independent of electron doping for $\omega >3\mid t\mid$. This
reflects an increase in the mobile carrier density, and indicates
that the spectral weight of the midinfrared sideband is taken from
the Drude absorption, then the spectral weight from both low
energy peak and midinfrared band represent the actual free-carrier
density. For a better understanding of the optical properties of
electron doped Mott insulators on a triangular lattice, we have
studied the conductivity at different temperatures, and the
results at electron doping $\delta=0.35$ for $t/J=-2.5$ with
temperature $T=0$ (solid line), $T=0.3J$ (dashed line), and
$T=0.7J$ (dotted line) are plotted in Fig. 2. It is shown that the
conductivity spectrum is temperature-dependent for $\omega <3\mid
t\mid$, and almost temperature-independent for $\omega>3\mid
t\mid$. The low-energy peak broadens and decreases in the height
with increasing temperatures, and there is a tendency towards the
Drude-like behavior. At the same time, the midinfrared band is
severely suppressed with increasing temperatures, and vanishes at
high temperatures, in qualitative agreement with these of hole
doped Mott insulators on a square lattice
\cite{tanner,feng1,feng2,stephan}. As in doped cuprates, the
charge transport is governed by the dressed fermion scattering,
therefore $\delta$ dressed fermion are responsible for the
conductivity, i.e., the optical conductivity in electron doped
cobaltates is carried by $\delta$ electrons. Since the strong
electron correlation is common for both hole doped cuprates and
electron doped cobaltates, these similar behaviors observed from
the optical conductivity are expected.

\begin{figure}[prb]
\epsfxsize=3.0in\centerline{\epsffile{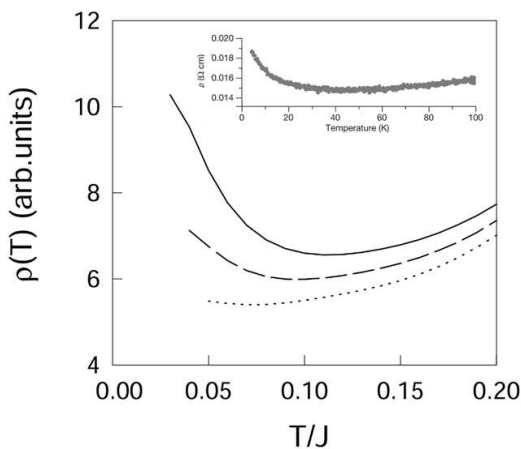}} \caption{The
resistivity of electron doped Mott insulators on a triangular
lattice as a function of temperature for $t/J=-2.5$ at electron
doping $\delta=0.36$ (solid line), $\delta=0.35$ (dashed line),
and $\delta=0.34$ (dotted line). Inset: the experimental result on
Na$_{0.35}$CoO$_{2}\cdot 1.3$H$_{2}$O taken from Ref. [1].}
\end{figure}

Now we turn to discuss the resistivity, which can be obtained in
terms of the conductivity as $\rho(T)=1/\lim_{\omega\rightarrow
0}\sigma (\omega)$, and the results for $t/J=-2.5$ at electron
doping $\delta=0.36$ (solid line), $\delta=0.35$ (dashed line),
and $\delta=0.34$ (dotted line) are plotted in Fig. 3, in
comparison with the experimental data \cite{takada} taken from
Na$_{x}$CoO$_{2}\cdot y$H$_{2}$O (inset). Our present results show
obviously that $\rho(T)$ is characterized by a crossover from the
high temperature metallic-like ($d\rho(T)/dT>0$) to low
temperature insulating-like behavior ($d\rho(T)/dT<0$), but the
metallic-like linear temperature dependence dominates over a wide
temperature range, which is qualitatively consistent with
experiments \cite{takada,wang}.

An explanation for this unusual charge transport can be found from
a competition between the kinetic energy and magnetic energy,
since the present $t$-$J$ model (4) is characterized by the
competition between the kinetic energy ($\delta t$) and magnetic
energy ($J$), with the magnetic energy $J$ favors the magnetic
order for spins, while the kinetic energy $\delta t$ favors
delocalization of electrons and tends to destroy the magnetic
order. Although both dressed fermions and spinons contribute to
the charge transport in the partial charge-spin separation
fermion-spin theory, the dressed fermion scattering dominates the
charge transport \cite{feng1,feng2}. The dressed fermion
scattering rate is obtained from the full dressed fermion Green's
function (then the dressed fermion self-energy and spectral
function) by considering the interaction between the dressed
fermion and spinon. In this case, the crossover from the high
temperature metallic-like to low temperature insulating-like
behavior in the resistivity of electron doped cobaltates is
closely related to this competition. In lower temperatures, the
dressed fermion kinetic energy is much smaller than the magnetic
energy, then the magnetic fluctuation is strong enough to severely
reduce the dressed fermion scattering and thus is responsible for
the insulating-like behavior in the resistivity. With increasing
temperatures, the dressed fermion kinetic energy is increased,
while the dressed spinon magnetic energy is decreased. In the
region where the dressed fermion kinetic energy is much larger
than the dressed spinon magnetic energy at higher temperatures,
the dressed fermion scattering would give rise to the
metallic-like resistivity.

We emphasize that this competition between the kinetic energy and
magnetic energy exists in all doped Mott insulators. However, in
the doped two-leg ladder Mott insulators, the charged carrier's
motion is also suppressed by interference effects between the two
legs \cite{feng5}, therefore in the doped two-leg ladder Mott
insulators both competition between the kinetic energy and
magnetic energy and interference effects between the two legs
cause the unusual charge transport. This is why even in the hole
doped two-leg ladder systems, the metallic-like state appears in
much higher doping concentration. On the other hand, hole doped
Mott insulators (opposite sign of hopping $t$) on a triangular
lattice have been discussed by many authors \cite{manuel,feng7},
where the conventional quasiparticle picture may be broken by the
effects of geometric spin fluctuation \cite{manuel}. These and
related differences \cite{feng7} between hole and electron doped
Mott insulators on a triangular lattice reflect that there is no
particle-hole symmetry, and the sign of $t$ is important
\cite{baskaran}.

In summary, we have studied the charge transport of electron doped
cobaltates within the $t$-$J$ model based on the partial
charge-spin separation fermion-spin theory. It is shown that the
conductivity spectrum shows a low-energy peak and a broad
midinfrared band, while the resistivity is characterized by a
crossover from the high temperature metallic-like to low
temperature insulating-like behavior. Our results also show that
the mechanism that causes this unusual charge transport is closely
related to the competition between the kinetic energy and magnetic
energy in the system.

Recent experimental studies \cite{kanigel} have shown that the
superconducting transition temperature in Na$_{x}$CoO$_{2}\cdot
y$H$_{2}$O is proportional to the electron doping concentration in
the underdoped regime, and satisfies the Uemura relation for hole
doped cuprates. This and other experiments have shown that there
is a remarkable resemblance in superconducting-state properties
between the electron doped cobaltate Na$_{x}$CoO$_{2}\cdot
y$H$_{2}$O and hole doped cuprates \cite{kanigel,schaak,takada},
manifesting that two systems have similar underlying
superconducting mechanism. Within the partial charge-spin
separation fermion-spin theory, the mechanism of superconductivity
in hole doped cuprates \cite{feng4} has been discussed, where
dressed holons interact occurring directly through the kinetic
energy by exchanging the dressed spinon excitations, leading to a
net attractive force between the dressed holons, then the electron
Cooper pairs originating from the dressed holon pairing state are
due to the charge-spin recombination, and their condensation
reveals the superconducting ground-state. In this case, the
electron superconducting transition temperature is determined by
the dressed holon pair transition temperature, and is proportional
to the hole doping concentration in the underdoped regime, in
agreement with experiments. Since the strong electron correlation
is common for both hole doped cuprates and electron doped
cobaltates, then it is possible that superconductivity in electron
doped cobaltates is also driven by the kinetic energy as in hole
doped cuprates. Following the discussions in Ref. \cite{feng4}, we
have studied this issue \cite{feng6}, and found that the
superconducting transition temperature in electron doped
cobaltates is suppressed to a lower temperature due to the strong
frustration. These and related theoretical results will be
presented elsewhere.

After submitting this paper, we became aware of recent optical
measurements \cite{nlwang} on Na$_{0.7}$CoO$_{2}$ supporting our
theoretical results.

\acknowledgments

The authors would like to thank Dr. Jihong Qin, Dr. Tianxing Ma,
and Professor Y.J. Wang for the helpful discussions. This work was
supported by the National Natural Science Foundation of China
under Grant Nos. 10125415, 90103024, and 10347102, and the
National Science Council.


\begin{references}
\bibitem [*] {add} To whom correspondence should be addressed.
\end{references}
\end{document}